\documentstyle[multicol,aps,epsf,amssymb,url]{revtex}

\def\Pr   {{\rm Pr}}
\def\hmu  {h_\mu}
\def\EE   {{\bf E} }
\def\SI   {{\bf S} }

\def\TI   {{\bf T} }
\def\ST   {{\tau} }
\def\l2   {{\rm log}_2}

\begin{document}


\title{Synchronizing to Periodicity: \\
The Transient Information and Synchronization Time of Periodic
Sequences}  

\author{David P. Feldman}
\address{College of the Atlantic, 105 Eden St., Bar Harbor, ME 
04609\\and Santa Fe Institute, 1399 Hyde Park Road, Santa Fe, NM 87501\\
Electronic Address: dpf@santafe.edu}

\author{James P. Crutchfield}
\address{Santa Fe Institute, 1399 Hyde Park Road, Santa Fe, NM 87501\\
Electronic Address: chaos@santafe.edu}

\date{\today}
\maketitle

\bibliographystyle{unsrt}

\begin{abstract}
We analyze how difficult it is to synchronize to a periodic sequence
whose structure is known, when an observer is initially unaware of
the sequence's phase. We examine the transient information $\TI$,
a recently introduced information-theoretic quantity that measures the
uncertainty an observer experiences while synchronizing to a sequence.
We also consider the synchronization time $\ST$, which is the average
number of measurements required to infer the phase of a periodic signal.
We calculate $\TI$ and $\ST$ for all periodic sequences up to and
including period $23$. We show which sequences of a given period
have the maximum and minimum possible $\TI$ and $\ST$ values, develop
analytic expressions for the extreme values, and show that in these
cases the transient information is the product of the total phase
information and the synchronization time. Despite the latter result,
our analyses demonstrate that the transient information and
synchronization time capture different and complementary structural
properties of individual periodic sequences --- properties, moreover,
that are distinct from source entropy rate and mutual information
measures, such as the excess entropy. 

\begin{center}

Santa Fe Institute~Working Paper 02-08-043
\end{center}

\end{abstract}

\begin{multicols}{2}

\section{Introduction}

Imagine you are about to begin observing a sequence of events. You
know the sequence is periodic; you even know the particular pattern
that will repeat. However, you do not know what phase the sequence
is in. How many observations, on average, would you have to make
before you know with certainty the sequence's phase?  And, as you
obtain this certainty, how uncertain are you about the phase?  Will
the answers to these questions be the same for all sequences of a
given period or are there differences between such sequences?  What
structural properties of a sequence determine the difficulty of
synchronization?  Here, we answer these questions.  

A natural place to begin is with information theory, since it has long
been used to analyze the statistical properties of sequences
that arise in a variety of settings, including dynamical systems,
time-series analysis, statistical mechanics, signal processing, and
cryptography \cite{Shan62,Cove91}. One of the central quantities in
these analyses is the {\em entropy rate} $\hmu$, the long-time measure
of unpredictability of sequences produced by an information source.
Although useful and important for quantifying randomness,  $\hmu$ does
not capture a sequence's structural properties: its correlation,
memory, or statistical complexity. Fortunately, today there are scores
of measures of these latter properties. For example, on the
information theoretic side, an oft-used measure of memory is the {\em
excess entropy}\/ $\EE$ --- the time-averaged, ``all-point'' mutual
information between a sequence's past and future
\cite{Crut83a,Gras86,Crut01a,Lind88b,Li91}.    

For any sequence of period $P$, it is well known that the entropy rate
$h_\mu$ vanishes and the excess entropy $\EE = \log_2 P$, since a
periodic sequence is asymptotically predictable and since $\log_2 P$ 
bits of information are needed to store in which of the $P$ possible
phases a symbol in the sequence is.  Thus, all periodic sequences of
the same period have the same entropy rate and excess entropy, and so
these quantities are unable to capture structural differences between
distinct sequences of the same period. In particular, $\hmu$ and $\EE$
cannot help answer the questions posed above. 

We turn instead to a recently introduced measure, the transient
information $\TI$ \cite{Crut01a,Crut01b}, which captures
information-theoretic differences between sequences of a given
period.  We will also make use of the synchronization time $\ST$,
defined as the average number of observations needed to synchronize to
a periodic information source.  We shall see that there are indeed
significant differences in the synchronization properties of periodic
sequences.  Furthermore, while $\TI$ and $\tau$ are related for
extreme cases --- minimum and maximum $\TI$ and $\tau$ sequences ---
there is a wide range of $\TI$ and $\tau$ values for
periodic sequences of the same period. 

In the following section we review information sources, entropy rate,
and excess entropy.  In Sec.~\ref{Synchronization.Section} we define
$\TI$ and $\ST$ more carefully.  After a brief discussion of
combinatorics and symmetry types of periodic sequences in
Sec.~\ref{Symmetry.Section}, in Sec.~\ref{Methods.Section} we discuss
methods used to calculate $\TI$ and $\ST$.  In
Sec.~\ref{Empirical.Section} we present the results of exhaustively
calculating $\TI$ and $\ST$ up to period $23$.  In
Secs.~\ref{Min.Max.Section} and \ref{Relationships.Section}, we
investigate these results, developing analytic expressions for the minimum
and maximum $\TI$ and $\ST$ values of a given period, exploring
relationships and bounds between the two synchronization measures.
Finally, in Sec.~\ref{Conclusion} we summarize and interpret the
results, pointing out several applications to coordination in
multiagent systems.


\section{Entropy Rate and Excess Entropy}

\subsection{Information Sources}

We shall be concerned with a one-dimensional infinite sequence of
variables: 
\begin{equation}
\stackrel{\leftrightarrow}{S} \, \equiv \,\ldots S_{-2} S_{-1} S_0 S_1
  \ldots  \;. 
\end{equation}
Here, the $S_t$'s are random variables that range over a finite set
${\cal A}$ of alphabet symbols.  In general, ${\cal A} = \{ 0,1,
\ldots , k-1 \}$, although in the following, we will restrict
ourselves to binary sequences. We denote a block or {\em word} of
$L$ consecutive variables by $S^L \equiv S_1 \ldots S_L$.  We follow
the convention that a capital letter refers to a random variable,
while a lowercase letter denotes a particular value of that variable.
Thus, $s^L = s_0 s_1 \cdots s_{L-1}$, denotes a particular symbol
block of length $L$. 

We assume that the underlying {\em information source} is described
by a shift-invariant measure $\mu$ on infinite sequences
$\cdots s_{-2} s_{-1} s_0 s_1 s_2 \cdots; s_t \in \cal A$
\cite{Gray90a}. The measure $\mu$ induces a family of distributions, 
$\{ {\rm Pr}(s_{t+1} , \ldots , s_{t+L}): s_t \in \cal A \}$, where
${\rm Pr}(s_t)$ denotes the probability that at time $t$ the random
variable $S_t$ takes on the particular value $s_t \in \cal A$ and
${\rm Pr} (s_{t+1} , \ldots , s_{t+L})$ denotes the joint probability
over blocks of $L$ consecutive symbols.  We assume that the
source is stationary; ${\rm Pr}(s_{t+1},\ldots, s_{t+L})={\rm
Pr}(s_1, \ldots , s_L )$.   

We are interested in periodic sequences; a sequence is {\em periodic}
of period $P$ if $s_i = s_{i+P}$ for all $i$. The {\em prime period}
of the sequence is the smallest such $P$.  We can specify a periodic
sequence by giving the smallest word --- the {\em prime word} --- that
is exactly repeated. The prime word is aperiodic, necessarily.  For
example, $\ldots 1100110011001100 \ldots$ is period $8$ (and period
$16$ and so on), but has prime word $1100$ and prime period $4$.

\subsection{Entropy Growth and Entropy Rate}

Here, we give a brief review of the information-theoretic description  
of sequences, concentrating on those that are periodic. For more
detail about information-theoretic measures of uncertainty and
structure in the context of general one-dimensional random sequences,
see, e.g., Refs.~\cite{Crut01a,Ebel97b,Bial00a} and references
therein. 

The {\em total Shannon entropy} of length-$L$ blocks is defined by:
\begin{equation}
H(L) \, \equiv \, - \sum_{ s^L \in {\cal A}^L } \Pr (s^L) \l2 \Pr
(s^L) \;, 
\end{equation}
where $L > 0$. The sum is understood to run over all possible blocks
of $L$ consecutive symbols. The units of $H(L)$ are {\em bits}. The
entropy $H(L)$ measures the average uncertainty in identifying one
sequence in the set of length-$L$ sequences.  Equivalently, $H(L)$
tells us how many yes-no questions, on average, are needed to
determine the value of $S^L$.  Another way to state this is that
$H(L)$ sets a lower bound on the size of the code, measured in bits,
needed to encode successive outcomes of the random variable $S^L$.
For more on this coding interpretation of the Shannon entropy, see,
e.g., Ref.~\cite{Cove91}.

\begin{figure}[tbp]
\epsfxsize=3.1in
\begin{center}
\leavevmode
\epsffile{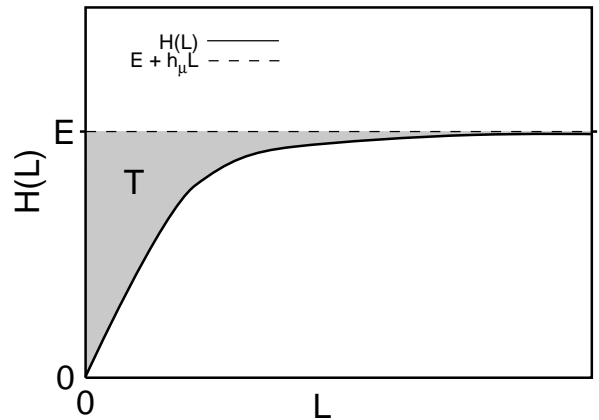}
\end{center}
\caption{Total Shannon entropy growth: A schematic plot of $H(L)$
  versus $L$. The entropy $H(L)$ of $L$-blocks increases monotonically
  and approaches the asymptote $\EE + h_\mu L$, where $\EE$ is the
  excess entropy and $h_\mu$ is the source entropy rate. The shaded area
  is the transient information $\TI$.
  }
\label{HvsL}
\end{figure}

The general behavior of $H(L)$ as a function of $L$ for periodic
information sources is shown schematically in Fig.~\ref{HvsL}.  Note
that we define $H(0) \equiv 0$. 

The {\em entropy rate} $\hmu$ is the rate of increase with
respect to $L$ of the total Shannon entropy in the large-$L$ limit:
\begin{equation}
    \hmu \equiv \lim_{L \rightarrow \infty} \frac{H(L)}{L} \; ,
\label{ent.def}
\end{equation}
where $\mu$ denotes the measure over infinite sequences that induces 
the $L$-block joint distribution ${\rm Pr} (S^L)$; the units are
{\em bits per symbol}. 
One can also define a finite-$L$ approximation to $h_\mu$, 
\begin{eqnarray}
  h_\mu(L) \, &=& \, H(L) - H(L\!-\!1) \;, 
\label{h.def} \\
 \, &=& \,  H[S_L | S_{L-1} S_{L-2} \ldots S_0 ]\;,
\label{h.condent}
\end{eqnarray}
where $H[X|Y]$ is the entropy of the random variable $X$ conditioned
on the random variable $Y$: $H[X|Y] \, \equiv \, - \sum_{x, y} {\rm
Pr}(x,y) \log_2 {\rm Pr(x|y)}$. One can then show \cite{Cove91} that 
\begin{equation}
h_\mu = \lim_{L \rightarrow \infty} h_\mu(L) \;.  
\label{h.alt}
\end{equation}
Note that Eq.~(\ref{h.def}) shows that $h_\mu(L)$ may be viewed as the
slope of $H(L)$.  For more on this point of view, see
Ref.~\cite{Crut01a}.  

\begin{figure}[tbp]
\epsfxsize=3.1in
\begin{center}
\leavevmode
\epsffile{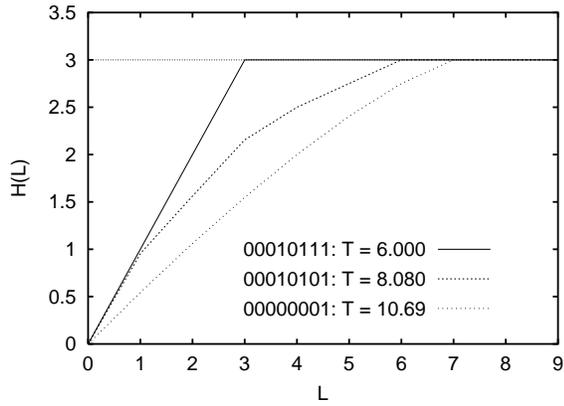}
\end{center}
\caption{Shannon entropy growth curves $H(L)$ for three different
period-$8$   sequences: $00010111$, $00010101$, and $00000001$.}
\label{Per8.HvsL}
\end{figure}

Eq.~(\ref{h.alt}) gives us another interpretation of $h_\mu$.  The
entropy rate quantifies the irreducible randomness or unpredictability
of the sequences; $h_\mu$ measures the randomness that persists even
after the statistics of longer and longer blocks are taken into
account.  Since all periodic sequences are, ultimately, predictable,
the entropy rate for all periodic sequences is $\hmu = 0$.  For a
period-$P$ sequence $h_\mu(P) = 0$ and $\hmu (L) \geq 0$, for $L =
1,2, \ldots, P\!-\!1$. This is illustrated in Fig.~\ref{Per8.HvsL} for
three different period-$8$ sequences; when the slope of $H(L)$
vanishes, $\hmu(L) = 0$.  

\subsection{Excess Entropy: Apparent Memory}

The entropy rate measures the randomness of sequences.  A
complementary quantity is the excess entropy $\EE$, which measures the
deviations of finite-$L$ estimates of $h_\mu$ from its asymptotic value:
\begin{equation}
\EE \, \equiv \, \sum_{L=1}^\infty [\hmu(L) - \hmu] \;,
\label{E.def}
\end{equation}
The units of $\EE$ are {\em bits}.  Using this definition, one can
show that the excess entropy is the subextensive part of $H(L)$
(see, e.g., \cite{Gras86,Crut01a,Bial00a,Shaw84,Neme00a,Li91}); that 
is, 
\begin{equation}
  \EE = \lim_{L \rightarrow \infty} [ H(L) - \hmu L ]\;.
\label{EEfromEntropyGrowth}
\end{equation}
This establishes a geometric interpretation for $\EE$, as shown in
Fig.~\ref{HvsL}; it is the $y$-intercept of the line to which $H(L)$
asymptotes. That is, when $\EE$ is finite, $H(L)$ asymptotes to
$\EE + \hmu L$.

Another way to understand excess entropy is through its expression as a
mutual information: the mutual information between the past and future
semi-infinite halves of the chain of random variables: 
\begin{equation}
  \EE \, = \,\lim_{L \rightarrow \infty}
  I[S_{-L+1}, S_{-L+2}, \cdots S_0 ; S_1 S_2 \cdots S_L] \;,
\label{EandMI}
\end{equation}
when the limit exists \cite{Crut01a,Li91}.  Eq.~(\ref{EandMI}) says
that $\EE$ measures the amount of historical information stored in the
present that is communicated to the future. 

Our focus here is on periodic sequences, for which it is easy to
show from one or another of the above definitions that
\begin{equation}
\EE = \l2 P ~.
\end{equation}
That is, the excess entropy is the total phase information stored
in the periodic source.  As noted above, the entropy of a random
variable gives the average number of yes-no questions (i.e., bits),
needed to determine the outcome of the variable. Since the observer
starts out equally ignorant of the initial phase, it follows that
$\EE$ is a lower bound on the amount of information (in bits) an
observer must extract in order to know in which phase the source
is --- that is, in order for it to be synchronized to the source. 

The entropy growth curves $H(L)$ for three different period-$8$
sequences are shown in Fig.~\ref{Per8.HvsL}.  Note that all curves
asymptote to the same line, ${\EE} + h_\mu L$.  Since $h_\mu = 0$
(periodic sequences) and ${\EE} = \log_2 P = \log_2 8 = 3$, the $H(L)$
curves asymptote to the horizontal line $H(L) = 3$ bits, when $L \geq
8$.  

Despite the fact that these three period-$8$ sequences have the same
$\hmu$ and ${\EE}$, Fig.~\ref{Per8.HvsL} shows that their entropy
growth curves $H(L)$ are not the same; for example, they reach the
linear asymptote at different $L$'s.  Do all period-$8$ curves have
different $H(L)$ behaviors? And do these different $H(L)$ behaviors
matter, or do $\hmu$ and $\EE$ suffice to characterize periodic
sequences? 

\section{Measures of Synchronization}
\label{Synchronization.Section}

\subsection{Transient Information}
\label{T_Section}

To address these questions, we consider the {\em transient information}
$\TI$, which was introduced by us in Ref.~\cite{Crut01a} and is defined
by summing the deviations of $H(L)$ from its linear asymptote: 
\begin{equation}
\TI \, \equiv \, \sum_{L=0}^\infty \left[ \EE + \hmu L - H(L) \right] \;. 
\label{T.def}
\end{equation}
Note that the units of $\TI$ are {\em bits $\times$ symbols} and that
$\TI \geq \EE$. The transient information is the shaded area in
Fig.~\ref{HvsL}.  We refer to $\TI$ as {\em transient}, since, unlike
$\hmu$ and $\EE$, it is dominated by nonasymptotic quantities: how
$H(L)$ behaves before it reaches the linear scaling form
${\bf E} + h_\mu L$.  By inspection, one sees that the three
period-$8$ sequences in Fig.~\ref{Per8.HvsL} all have different
$\TI$ values.

As noted above, for finite-$\EE$ processes $H(L)$ scales as
$\EE + h_\mu L$ for large $L$ \cite{Crut01a}. When this scaling form is
attained, we say that the observer is {\em synchronized} to the source.
In other words, when
\begin{equation}
\TI (L) \, \equiv \, \EE + \hmu L - H(L) \, = \, 0 \;,
\label{OSSyncCondition}
\end{equation}
we say the observer is synchronized at length-$L$ sequences.  The
quantity $\TI (L)$ provides a measure of the departure from 
synchronization. Note that $\TI(L)$ is non-negative. Looking at
Eq.~(\ref{T.def}), we see that the transient information $\TI$ may be
written as a sum of the $\TI(L)$'s:
\begin{equation}
 {\bf T} \, = \, \sum_{L=0}^\infty {\bf T}(L) \;. 
\end{equation}

To give the transient information a more precise interpretation, let
us consider in more detail the synchronization scenario we have
in mind.  An observer begins making measurements of a sequence, seeing
one symbol at a time.  The observer knows the periodic sequence it is
about to start seeing, but it doesn't know the phase. That is, the
observer knows the period and the prime word that will be repeated.
The task for the observer is to make measurements and determine the
sequence's phase. Exact prediction is possible from this point onwards,
though {\em not} before.  How uncertain is the observer during this
synchronization process?

To answer this question, we must introduce some additional notation.
Before the observer is synchronized, its knowledge is characterized by
a distribution over the $P$ possible phases of the periodic sequence.
Equivalently, the sequence's phases may be viewed as the periodic
source's internal states.  Let $\varphi$ denote a particular phase, 
and let $\Phi$ denote the set of all phases.  Clearly, $|\Phi| =
P$.  Next, let ${\rm Pr}( \varphi | s^L )$  denote the probability, as
inferred by the observer, that the sequence is in phase $\varphi$,
given that it has just seen the particular sequence of symbols
$s^L$.  The entropy of the distribution of $\varphi$ measures the
observer's average uncertainty of the phase $\Phi$.  Averaging this
uncertainty over the possible length-$L$ observations, we obtain the
{\em average state-uncertainty}: 
\begin{equation}
  {\cal H}(L) \, \equiv \,
  - \sum_{s^L} {\rm Pr}( s^L )\sum_{\varphi \in \Phi } 
  {\rm Pr}( \varphi | s^L) \log_2  {\rm Pr}( \varphi | s^L) \;.
\label{script.H.def}
\end{equation}
The quantity ${\cal H}(L)$ can be used as a criterion for
synchronization. The observer is synchronized to the sequence when
${\cal H}(L) = 0$ --- that is, when it is completely certain about the 
phase, or internal state, $\varphi \in \Phi$ of the source generating
the sequence. And so, when the condition in Eq.~(\ref{OSSyncCondition})
is met, we see that ${\cal H}(L) = 0$, and the uncertainty associated
with the next observation is $0$. 

However, while the observer is still unsynchronized ${\cal H}(L) >
0$.  The average total uncertainty experienced by the
observer during the synchronization process is the total {\em
synchronization information} $\SI$: 
\begin{equation}
\SI \equiv \sum_{L=1}^P {\cal H}(L) \;.
\label{SI.def}
\end{equation}
The synchronization information measures the total uncertainty ${\cal
H}(L)$ experienced by an observer during synchronization.  For the
periodic sequences under consideration here, it turns out that  
\begin{equation}
\SI \, = \,\TI  \;.
\label{SyncTheoremPeriodic}
\end{equation}
That is, the transient information $\TI$ is equal to the total
synchronization information $\SI$.  

Eq.~(\ref{SyncTheoremPeriodic}) is a special case of a theorem
recently proved by us in Ref.~\cite{Crut01a} and discussed further in
Ref.~\cite{Crut01b}.  Here, we briefly sketch the main argument behind
Eq.~(\ref{SyncTheoremPeriodic}).  At $L=0$, no measurements have been
made and the observer posits that each phase is equally likely.
The state-uncertainty is thus $\log_2 P = \EE$.  After $L$
observations have been made, the observer has gained, on average, 
$H(L)$ bits of information about the phase of the process.  As a
result, the average state-uncertainty ${\cal H}(L)$ is now $\EE -
H(L)$.  Plugging this last observation into Eq.~(\ref{SI.def}),
Eq.~(\ref{SyncTheoremPeriodic}) follows from the definition of $\TI$,
Eq.~(\ref{T.def}). 

As a result of Eq.~(\ref{SyncTheoremPeriodic}), the transient
information $\TI$ provides a direct measure of how difficult or
confusing it is to synchronize to a periodic sequence.  If a periodic 
sequence has a large $\TI$, then on average an observer will be
highly uncertain about its phase during the synchronization process.
The transient information measures a structural property of a sequence
--- a property captured neither by the entropy rate nor by the excess
entropy.  If the observer is in a position where it must take
immediate action, it does not have the option of waiting for full
synchronization. In this circumstance, the synchronization information 
$\SI$ provides an information-theoretic average-case measure of the
error incurred by the observer during the synchronization process.  

\subsection{Synchronization Time}
\label{tau.section}

In addition to measuring the total uncertainty experienced during
synchronization, we can ask a related question: On average, {\em how
many}\/ measurements must be made before the observer is certain in
which state the source is?  We call the number $\ST$ of
measurements, averaged over the $P$ possible starting phases, the
{\em synchronization time}. 

For example, consider the periodic sequence $(0001)^\infty$. There
are four possible phases in which one might begin to observe this
sequence. If the first symbols parsed are $1-0-0-0$, then the observer
is synchronized after the first observation (a $1$). If the observer 
initially sees $0-1-0-0$, it is synchronized after two symbols. If the
observer sees either $0-0-1-0$ or $0-0-0-1$, it is synchronized after
three measurements. Each of these initial measurement sequences is
equally likely. Thus, on average, it will take 
\begin{equation}
  \ST \, = \, \frac{1 + 2 + 3 + 3}{4} \, = \, 2.25
\end{equation}
measurements before the observer is synchronized to $(0001)^\infty$.  

Below, we conduct a survey of the transient information $\TI$ and
synchronization time $\ST$ of periodic sequences. Although there
is an overall scaling between them, perhaps somewhat surprisingly,
we shall see that $\TI$ and $\ST$ capture fundamentally different
properties of a sequence. We shall also see that within different
sequences of a given period there are wide variations of $\TI$ and
$\ST$.  First, however, we pause to consider some combinatorial
properties and symmetry classes of periodic sequences.    

\section{Entropic Symmetry Types and Combinatorics of Periodic
Sequences} 
\label{Symmetry.Section}

How many periodic sequences of a given prime period are there?
Aperiodic length-$P$ words that are related by an application of a
cyclic permutation group $C_P$ give the same periodic sequence. 
Thus, for our purposes --- considering infinitely repeated periodic
sequences --- two length-$P$ words related via the group operation
$C_P$ are equivalent; they yield identical infinite sequences.  For
example, the aperiodic words $001$, $010$, and $100$ all yield the
same, infinite, period-$3$ sequence. 


However, we are interested in looking for distinct $H(L)$ versus $L$
behaviors, and this induces another set of equivalences on the
period-$P$ sequences.  Because $H[X]$ depends only on the distribution
of $X$, and not on the values $X$ assumes, the entropy of a random
variable is unchanged under a one-to-one mapping of alphabet
symbols \cite{Cove91}. For the binary cases we are interested in here,
this means that the mapping $0 \leftrightarrow 1$ leaves $H(L)$
unchanged; for our purposes, the sequence $(011)^\infty$ is identical to
$(100)^\infty$. In terms of group action, swapping $1$ and $0$ is the
action of the symmetric group $S_1$. So, any two sequences that are
based on binary aperiodic length-$P$ words related to each other by
some combination of $S_1$ and $C_P$ will have identical $H(L)$ versus
$L$ behavior. 

The number $L_2$ of aperiodic binary words inequivalent under
$S_1 \times C_P$ is given by:
\begin{equation}
  L_2(P) \, = \, \frac{1}{P} \sum_{d|P, d \;{\rm odd} } \mu( d )
  2^{P/d} \;.  
\label{periodic.words}
\end{equation}
The sum runs over all odd positive divisors $d$ of $n$, and $\mu(m)$
is the M\"obius function: $\mu(m) = 0$, if $m$ is the product of
nondistinct primes; $+1$, if $m$ is the product of an even number of
distinct primes; and $-1$, otherwise.
This result was first obtained by Fine \cite{Fine58a} in 1958 and
simplified a few years later by Gilbert and Riordan \cite{Gilb61a}.
These words are also related to a version of the famous necklace
problem from combinatorics:  the words inequivalent under $S_1 \times 
C_P$ are exactly those $2$-color necklace sequences with prime period
$P$ where interchanging bead color is allowed, but turning the
necklace over is not.  

These aperiodic words are also related to the so-called Lyndon words.
A {\em Lyndon word} is aperiodic and the lexicographically least among
its rotations.  An {\em unlabeled Lyndon word} is aperiodic and
lexicographically least among its rotations and
relabelings. Eq.~(\ref{periodic.words}), then, also counts the number
of binary unlabeled Lyndon words. For more discussion, see
Ref.~\cite{Cos} and sequence \#A000048 of Ref.~\cite{Sloa}. Titsworth
\cite{Tits64b} has pointed out that Eq.~(\ref{periodic.words}) also
gives the number of distinct finite-state machines needed to produce
all periodic binary sequences of prime period $P$.

\section{Methods}
\label{Methods.Section}

We calculated the transient information $\TI$ and synchronization time
$\ST$ for all distinct periodic sequences (unlabeled Lyndon words) up
to and including period $P = 23$. To enumerate all of these words of a
given period, we used the efficient algorithms available at
Ref.~\cite{Cos}. 

\begin{figure}[tbp]
\epsfxsize=2in
\begin{center}
\epsffile{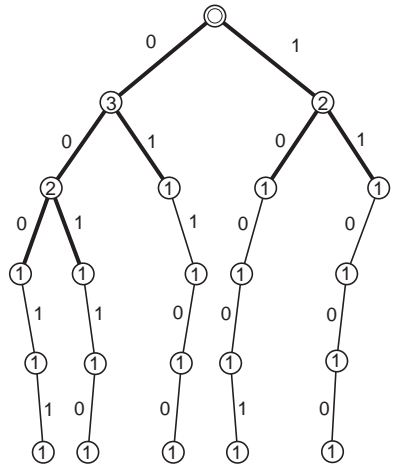}
\end{center}
\caption{Parse tree for the period-$5$ sequence with prime
  word $00011$. Unvisited paths are not shown. The synchronizing
  words are shown with bold paths from the start node.}
\label{Per5.Tree}
\end{figure}

\begin{table}[tbp]
\begin{center}
\begin{tabular}{c c c }
Prime Word & $\TI$ & $\ST$   \\
\hline
\hline
000001 & 6.97905 & 3.33333 \\
000011 & 5.37744 & 2.83333 \\
000101 & 5.58496 & 3.33333 \\
000111 & 4.83659 & 2.66667 \\
001011 & 4.83659 & 2.66667 \\
\end{tabular}
\end{center}
\caption{The transient information $\TI$ and synchronization time
$\ST$ of all five distinct period-$6$ sequences.  }
\label{Period.6.Table}
\end{table}

To calculate $\TI$ and $\ST$ for a given length-$P$ word, we begin by
determining the frequency of occurrence of all its subwords. This is
done by parsing the word and its $P-1$ cyclic permutations into a
tree, whose paths from the root to any node are the subwords and each
of whose nodes contain the number of subwords found that lead to it
from the root. 

An example is shown in Fig.~\ref{Per5.Tree}. Consider the sequence
whose prime word is the length-$5$ word $00011$. We start at the tree 
root, indicated by the double circle at the top, and follow the leaves
labeled with the appropriate symbol. Each time we cross a node we
increment the count there by one. We repeat this procedure for all $5$
cyclic permutations of $00011$. 

Once the parse tree is built, $\TI$ can be easily and exactly
calculated. First, the node counts are turned into probabilities by
normalizing --- dividing each count by the period $P$.  Reading across
the tree at level $\ell$ gives the probability of subwords of length
$\ell$.  In the $P=5$ example considered here, ${\rm Pr}(00) = 2/5$,
${\rm Pr}(01) = 1/5$, ${\rm Pr}(10) = 1/5$, and ${\rm Pr}(11) = 1/5$.
From these probabilities, the block entropies $H(L)$, and, in turn,
$\TI$, follow directly by using Eq.~(\ref{T.def}) and noting that
$\EE = \log_2 P$ and $\hmu = 0$. 

Calculating $\ST$ is only a bit more involved.  After building the
parse tree, we reparse each cyclic permutation of the word.  As we
reparse and proceed down the tree, we monitor the node counts.  For
each cyclic permutation of the prime word, we follow the corresponding
path from the root.  When we come to the first node that has a count
of $1$, we are synchronized.  That is, for the $i^{\rm th}$ cyclic
permutation of the word, the level $\ell$ at which we first encounter
a node count of $1$ is the synchronization time $\ST_i$ for that
permutation.  The average synchronization time $\ST$ is then simply: 
\begin{equation}
    \ST \, = \,\frac{1}{P} \sum_{i=0}^{P-1} \ST_i \;. 
\end{equation}
For the example in Fig.~\ref{Per5.Tree}, we have
\begin{equation}
  \ST \, = \, \frac{1}{5} (3 + 3 + 2 + 2 + 2) \, = \, 2.4 \;. 
\end{equation}
As for $\TI$, the above method yields an exact value for $\ST$. 

For example, in Table \ref{Period.6.Table} we show the results of
calculating the transient information $\TI$ and the synchronization
time $\ST$ for all five distinct period-$6$ sequences.  (The reader
may find it helpful to verify these results.)

\section{Empirical Results}
\label{Empirical.Section}

\begin{figure}[tbp]
\epsfxsize=3.1in
\begin{center}
\leavevmode
\epsffile{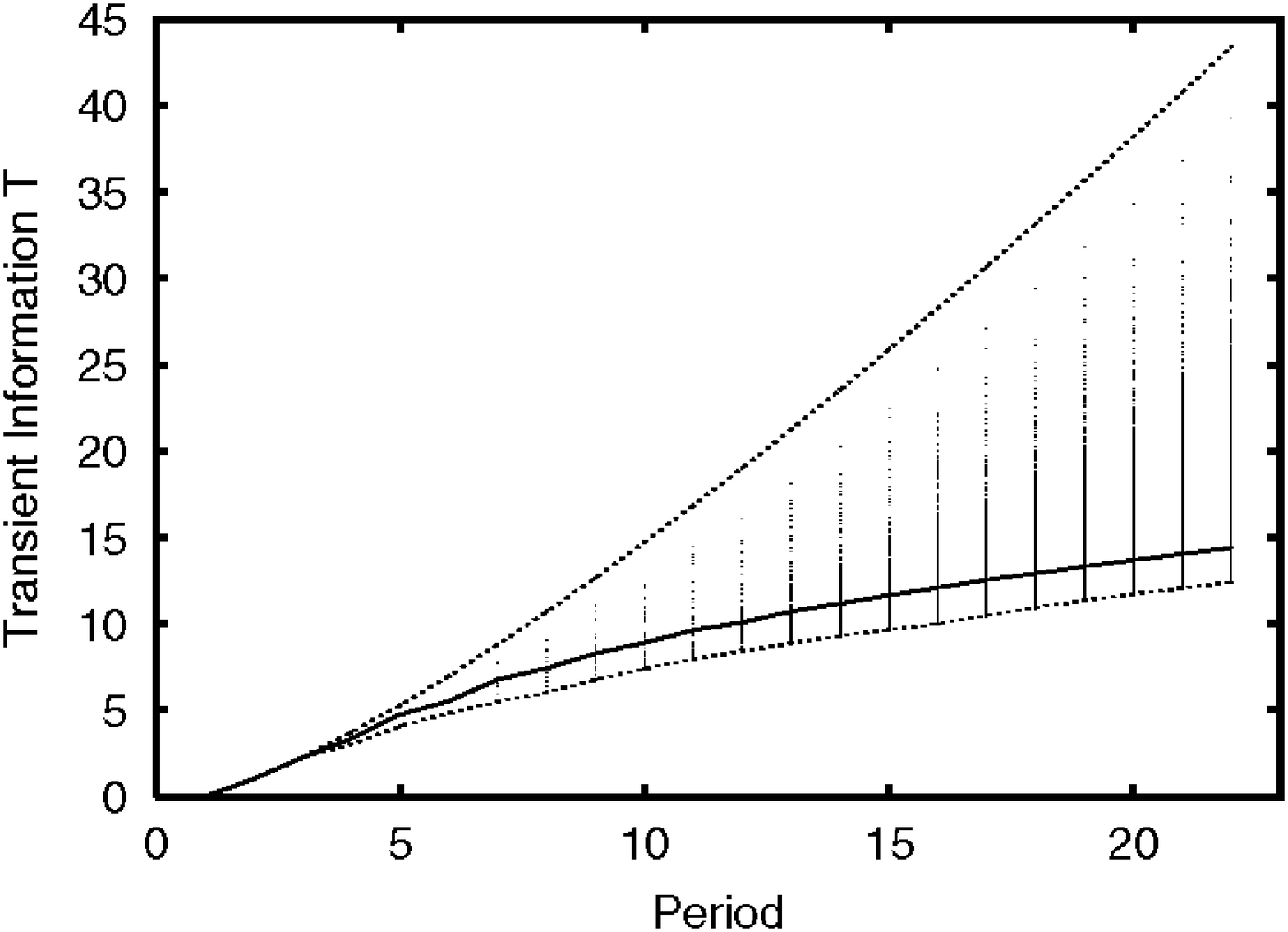}
\end{center}
\caption{Scatter plot of the transient information $\TI$ for all
distinct periodic sequences as a function of their prime period $P$,
up to $P = 22$. The dashed lines are the maximum and minimum $\TI$
values computed analytically; see text. The solid line gives the
estimated average $\TI$ value at the given period.}
\label{TvsL}
\end{figure}

\begin{figure}[tbp]
\epsfxsize=3.1in
\begin{center}
\leavevmode
\epsffile{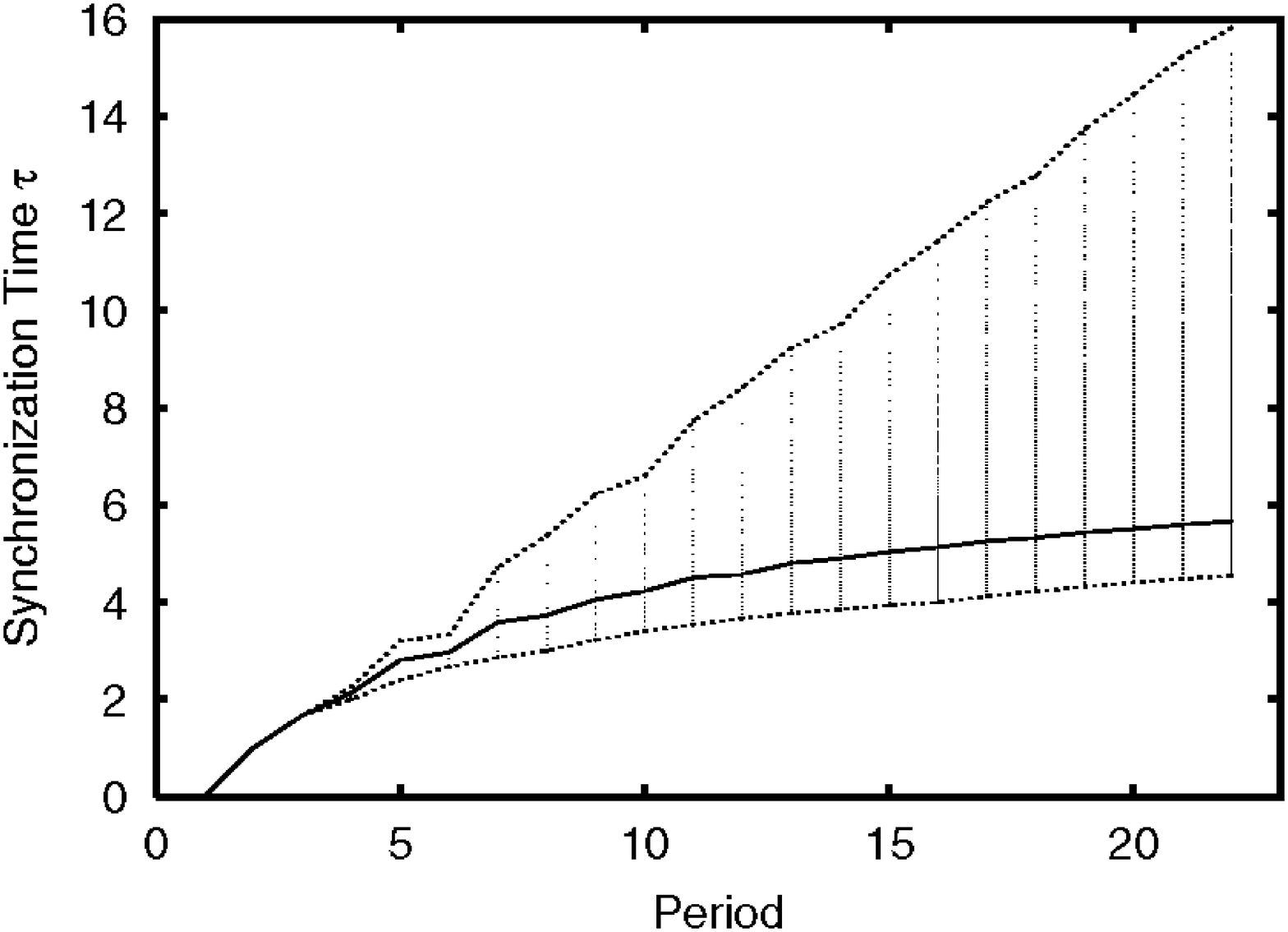}
\end{center}
\caption{Scatter plot of the synchronization time $\ST$ for all
distinct periodic sequences as a function of their period $P$,
up to $P = 22$. The dashed lines are the maximum and minimum
$\ST$ values computed analytically; see text. The solid line gives
the estimated average $\ST$ value at the given period.}
\label{TauvsL}
\end{figure}

\begin{figure}[tbp]
\epsfxsize=3.4in
\begin{center}
\leavevmode
\epsffile{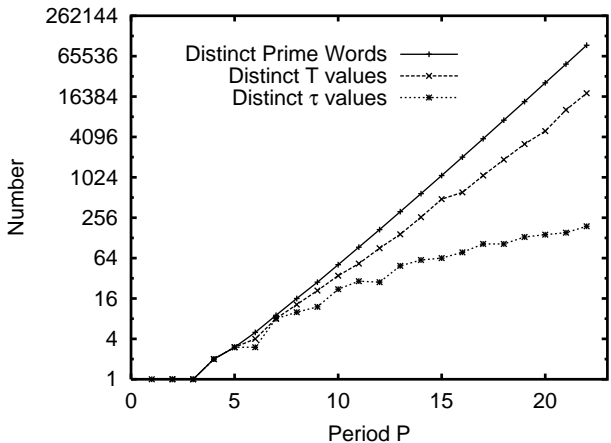}
\end{center}
\caption{The number of distinct periodic sequences, distinct $\TI$
  values, and distinct $\ST$ values as a function of period $P$. Note
  the logarithmic scale on the vertical axis.} 
\label{numbers}
\end{figure}

\begin{figure}[tbp]
\epsfxsize=3.4in
\begin{center}
\leavevmode
\epsffile{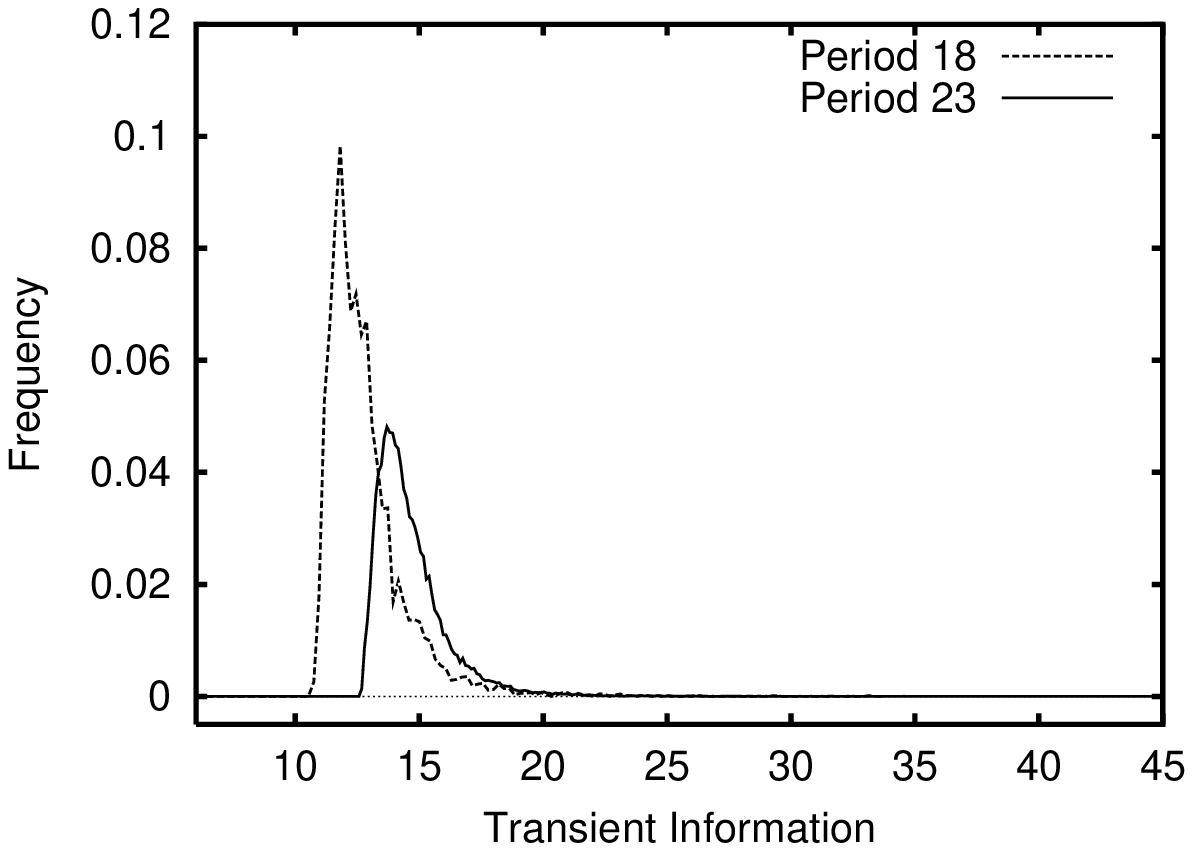}
\end{center}
\caption{Normalized histogram of the distribution of $\TI$ values for
  period-$18$ and period-$23$ sequences.}
\label{THistogram}
\end{figure}

\begin{figure}[tbp]
\epsfxsize=3.4in
\begin{center}
\leavevmode
\epsffile{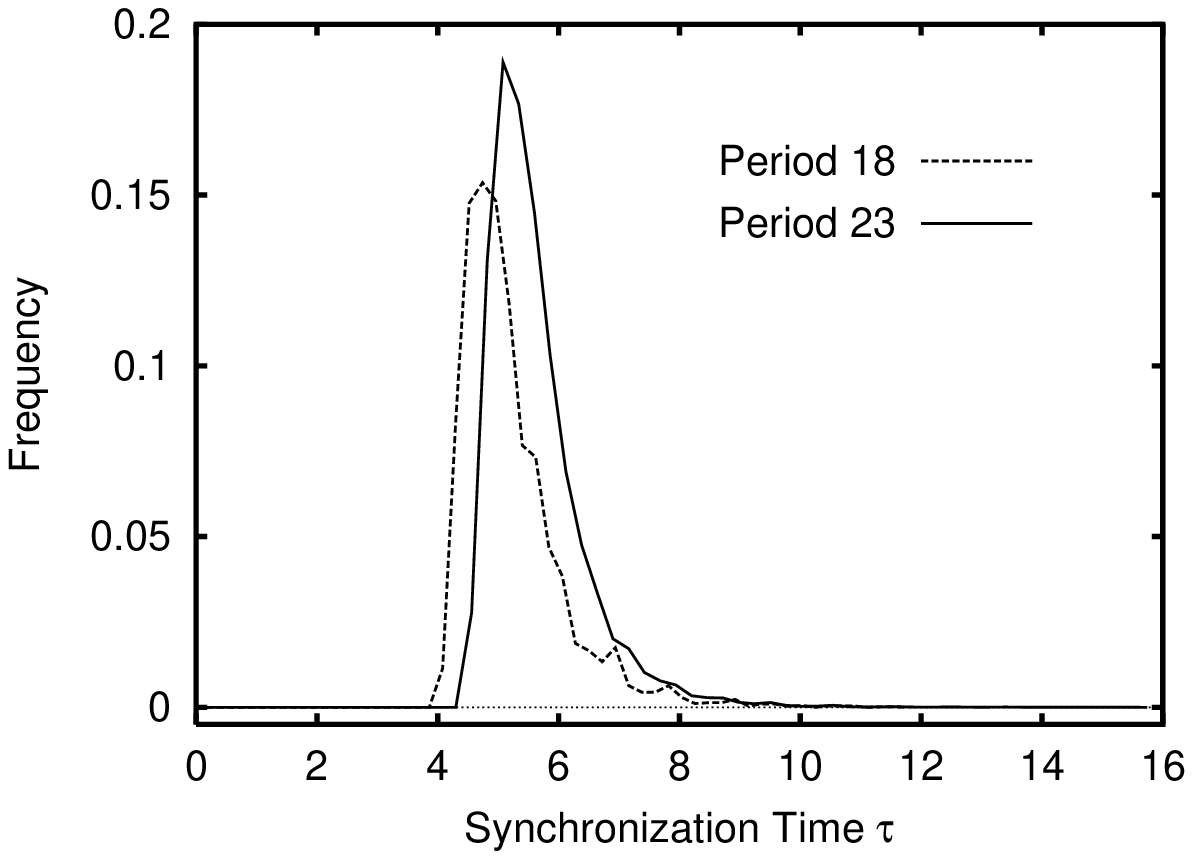}
\end{center}
\caption{Normalized histogram of the distribution of $\ST$ values for
  period-$18$ and period-$23$ sequences.}
\label{TauHistogram}
\end{figure}

We empirically investigated the behavior of $\TI$ and $\ST$ by first
exhaustively enumerating all distinct periodic sequences up to and
including period $P = 23$ and then exactly calculating these two
quantities for each using the methods of the previous section. The
results are summarized in Figs.~\ref{TvsL}, \ref{TauvsL},
\ref{numbers}, \ref{THistogram}, and \ref{TauHistogram}.  In Table
\ref{BigTable} we show, for periods $2$ through $23$, the number of
distinct periodic sequences, the number of distinct $\TI$ and $\ST$
values, and the mean, minimum, and maximum $\TI$ and $\ST$ values.

Based on these results, one can make a host of observations. First,
note that $\TI$ and $\ST$ are {\em not} the same for all sequences of
a given period. This can be seen in our period-$6$ results of Table
\ref{Period.6.Table}, as well as in Figs.~\ref{TvsL}, \ref{TauvsL}, and
\ref{numbers}.  Thus, within a given period, there are many different
synchronization behaviors. As noted before, these differences are not
accounted for by the entropy rate $h_\mu$ and the excess entropy
$\EE$, since $\EE$ and $h_\mu$ are the same for all sequences with
the same period.

\end{multicols}

\begin{table}
\begin{center}
\begin{tabular}{r r r c c c r c c c }
Period & Distinct & Distinct
& $\langle \TI \rangle$ & $\TI_{\rm min}$ & $\TI_{\rm max}$
& Distinct & $\langle \ST \rangle$ & $\ST_{\rm min}$ & $\ST_{\rm max}$ \\
$P$ & Periodic Sequences & $\TI$ Values & & & & $\ST$ Values & & &\\
\hline
\hline
3&	1&	1&	2.25163&	2.25163&	2.25163&	    1&	1.66667&	1.66667&	1.66667\\
4&	2&	2&	3.34436&	3.00000&	3.68872&	2&	2.12500&	2.00000&	2.25000\\	
5&	3&	3&	4.73957&	4.07291&	5.27291&	3&	2.80000&	2.40000&	3.20000\\	
6&	5&	4&	5.52293&	4.83659&	6.97905&	3&	2.96667&	2.66667&	3.33333\\	
7&	9&	8&	6.77183&	5.48662&	8.78940&	8&	3.58730&	2.85714&	4.71429\\	
8&	16&	13&	7.42624&	6.00000&	10.6907&	10&	3.72656&	3.00000&	5.37500\\	
9&	28&	21&	8.27733&	6.76598&	12.6728&	12&	4.05556&	3.22222&	6.22222\\	
10&	51&	35&	8.89152&	7.39483&	14.7274&	22&	4.22549&	3.40000&	6.60000\\	
11&	93&	53&	9.63275&	7.94889&	16.8480&	29&	4.50733&	3.54545&	7.72727\\	
12&	170&	90&	10.0802&	8.42155&	19.0290&	28&	4.57157&	3.66667&	8.41667\\	
13&	315&	145&	10.7162&	8.88705&	21.2656& 	49&	4.80293&	3.76923&	9.23077\\	
14&	585&	261&	11.1637&	9.29398&	23.5540&	60&	4.89634&	3.85714&	9.71429\\	
15&	1091&	484&	11.6530&	9.66732&	25.8906&	64&	5.03367&	3.93333&	10.7333\\	
16&	2048&	610&	12.0846&	10.0000&	28.2725&	78&	5.13293&	4.00000&	11.4375\\	
17&	3855&	1091&	12.5298&	10.4930&	30.6969&	104&	5.25312&	4.11765&	12.2353\\	
18&	7280&	1878&	12.9133&	10.9359&	33.1614&	104&	5.33134&	4.22222&	12.7778\\	
19&	13797&	3205&	13.3158&	11.3449&	35.6640&	132&	5.43299&	4.31579&	13.7368\\	
20&	26214&	5015&	13.6772&	11.7168&	38.2027&	143&	5.50798&	4.40000&	14.4500\\	
21&	49929&	10355&	14.0405&	12.0774&	40.7759&	153&	5.59134&	4.47619&	15.2381\\	
22&	95325&	16031&	14.3815&	12.4083&	43.3818&	191&	5.66223&	4.54545& 15.8182\\	
23 & 182361 & 27322 & 14.7179 & 12.7191 & 46.0192 & 207 & 5.73570 & 4.60870
& 16.7391 \\ 
\end{tabular}
\caption{The number of distinct periodic sequences, the number of
  distinct $\TI$ and $\ST$ values, and the average, minimum, and
  maximum $\TI$ and $\ST$ values as a function of period, up to $P = 23$.}      
\label{BigTable}
\end{center}
\end{table}

\begin{multicols}{2}

Second, there are words with the same period that have identical
$\TI$ or identical $\tau$ values.  This can be seen in
Fig.~\ref{numbers}, in which we have plotted the count data shown in
Table \ref{BigTable}. The number of distinct $\TI$ and $\ST$ values
clearly grows less quickly than the total number of distinct periodic
sequences.  Note also that the number of distinct $\TI$ and $\ST$
values grow at different rates.  This provides direct
evidence that the transient information and the synchronization time
measure different qualities.  For example, in Table \ref{Period.6.Table},
note that two sequences have the same $\ST$ but different $\TI$'s.
The converse also occurs, but much less frequently.  There are
sequences with the same $\TI$ but different $\ST$ values, but this
does not occur until $P = 12$.  As can be seen from both Table
\ref{BigTable} and Fig.~\ref{numbers}, $\ST$ tends to be substantially
more degenerate than $\TI$; $\ST$ is somehow a coarser measure of a
sequence's synchronization properties.  

Third, note that for a given period there is a considerable range of
transient information and synchronization time values.  This can be
seen in Figs.~\ref{TvsL} and \ref{TauvsL}, where the minimum and
maximum $\TI$ and $\ST$ values are shown as dashed lines.  This
suggests that there are significant differences in synchronization
behaviors among different sequences with the same period.  We
will return to the question of minimum and maximum $\TI$ and $\ST$
sequences in some detail in Sec.~\ref{Min.Max.Section}. 

Another way to view the range of $\TI$ values is found in
Fig.~\ref{THistogram}, in which we plot the distribution of $\TI$ values
for period-$16$ and period-$23$ sequences.  Note the asymmetry;
most $\TI$ values are closer to the minimum $\TI$ than the
maximum $\TI$.  Similar observations can be made about the
distribution of $\ST$ values shown in Fig.~\ref{TauHistogram}. 

Finally, note from Fig.~\ref{numbers} and Table \ref{BigTable}
that the number of distinct $\ST$ values is not strictly
monotonic with increasing periodicity.  Curiously, the number of
distinct $\ST$ values decreases as the period goes from $11$ to
$12$ and it remains constant in going from $P = 5$ to $P= 6$,
while it increases for all other $P$ increases.

\section{Minimum, Maximum, and Average $\TI$ and $\ST$}
\label{Min.Max.Section}

Here we develop analytical expressions for the extreme behaviors of
the transient information and the synchronization time, and we determine
how these extreme values grow asymptotically with period.  This
analysis allows us to derive a number of important conclusions about
which complementary structural features these quantities capture. 

First, some notational preliminaries.  For a given period, we denote
by $\langle \TI \rangle$ the average transient information, averaged
over all distinct prime words.  Similarly, $\TI_{\rm min}$ and
$\TI_{\rm max}$ denote the minimum and maximum $\TI$ values at a given
period, respectively.  We use similar notation for the average,
maximum, and minimum $\ST$ values. 

\subsection{Maximum-$\TI$ Sequence}

A sequence's transient information $\TI$ is large if $H(L)$ approaches
its asymptotic value of $\EE = \log_2 P$ slowly. This can be deduced
graphically from Figs.~\ref{HvsL} and \ref{Per8.HvsL}. The $H(L)$
curve that grows most slowly is that in which the distribution over
subwords is maximally nonuniform, since the entropy of a random variable
decreases as its distribution departs from uniformity.  For a
given period $P$, the sequence with the maximum value of $\TI$ 
is one consisting of $P\!-\!1$ $0$'s followed by an isolated $1$,
since this sequence is the one whose distribution over subwords is as
nonuniform as possible. This behavior is seen in one of the
period-$8$ sequences shown in Fig.~\ref{Per8.HvsL}. 

\begin{figure}[tbp]
\epsfxsize=3.2in
\begin{center}
\leavevmode
\epsffile{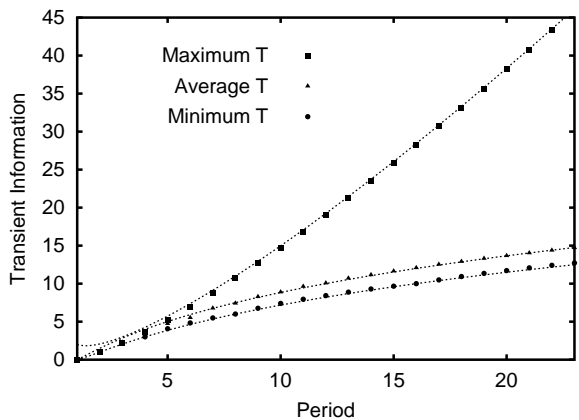}
\end{center}
\caption{The minimum, maximum, and average transient information
values plotted as a function of period.  The solid points are our exact
results; the dashed lines are the approximations developed in the text:
Eq.~(\ref{Tmax.approx}) for $\TI_{\rm max}$, Eq.~(\ref{Tmin.approx})
for $\TI_{\rm min}$, and Eq.~(\ref{Tave.fit}) for $\langle \TI \rangle$. }
\label{T.max.min}
\end{figure}

A direct calculation for the maximal-$\TI$ sequence yields:
\begin{equation}
 \TI_{\rm max} \, = \, \frac{1}{P} \sum_{n=2}^{P} n \log_2 n \;.
\label{Tmax.exact}
\end{equation}
To develop a functional form for $\TI_{\rm max}$ we approximate the
sum by an integral.  After some work, one obtains, 
\begin{eqnarray}
 \TI_{\rm max} \, & \approx & \, \frac{1}{\ln(16) P} \left[ \, 
     2 - P - P^2 + 4 \ln(2) \right. \nonumber \\ 
& & \;\; +  \; \left. P^2\ln(P) + (1\!+\!P)^2 \ln(1\!+\!P) \, \right] \;.
\label{Tmax.approx}
\end{eqnarray}
For large $P$, this scales as
\begin{equation}
 {\bf T}_{\rm max} \, \approx \, \frac{1}{2} P \, \l2 P \;.
\label{Tmax.Scaling}
\end{equation}
The values obtained from Eq.~(\ref{Tmax.exact}) agree with the exact
numerical results shown in Table \ref{BigTable}.  In Fig.~\ref{T.max.min}
we plot the approximation of Eq.~(\ref{Tmax.approx}) and compare them
with our exact numerical results.  As expected, the approximation is
quite good and improves at larger $P$. 

The synchronization time $\tau$ for the maximum-$\TI$ word can be
calculated directly by noticing that one is synchronized as soon as
one sees a $1$, or after $P-1$ measurements, whichever comes first.
One obtains, for large $P$: 
\begin{equation}
\tau_{{\rm max \; T}} \sim \frac{P-1}{2} \;,
\label{tau.for.Tmax}
\end{equation}

\subsection{Minimum-$\TI$ Sequences}

The minimum-$\TI$ word is particularly easy to analyze when the period
is a power of $2$.  We begin with this special case and use it to
derive a simple form for $\TI_{\rm min}$ as a function of the period.  

The sequences with the minimum transient information $\TI_{\rm min}$
correspond to those whose $H(L)$ curves rise most quickly to
$\EE$. This is achieved by a sequence whose distribution over
subwords, at each subword length, is most nearly uniform and so gives
high entropy: the measurements are most informative about the phase.

When the period $P = 2^M$, where $M$ is an integer, the distribution
over subwords for the $\TI_{\rm min}$ sequence is exactly uniform.
Words with this property are known as {\em de Bruijn} sequences.  A
binary de Bruijn sequence of order $k$ is a binary circular string
containing every binary substring of length $k$ exactly once. For
example, the lexicographically smallest order-$3$ binary de Bruijn
sequence is $00010111$.  Note that this was one of the period-$8$
sequences whose $H(L)$ curve was plotted in Fig.~\ref{Per8.HvsL}.  In
that figure, the rapid convergence of $H(L)$ to $\log_2 8 = 3$ is
clearly seen.   

Because every subword appears with equal probability, $H(L) = L$,
until $L = \log_2 P $.  That is, $H(L)$ converges linearly to the
excess entropy $\EE = \log_2 P$. In other words, the source appears
to be a fair coin until synchrony is achieved. At that time, exact
predictability becomes possible. This linear convergence can be seen
in Fig.~\ref{Per8.HvsL}.  The transient information is thus given by 
the area between $H(L)$ and $\log_2 P$;
\begin{equation}
 \TI_{\rm min} \, = \, \sum_{n=1}^{\log_2 P} n \;.
\end{equation}
Evaluating the sum, one has:
\begin{equation}
 \TI_{\rm min} \, \approx \, \frac{1}{2}\left(  \log_2^2 P + \log_2 P
 \right) \;. 
\label{Tmin.approx}
\end{equation}
Equivalently, since $\EE = \log_2 P$, we may write this as:
\begin{equation}
  \TI_{\rm min} \, \approx \, \frac{1}{2} \, \EE \,(\EE+1) \;. 
\label{Tmin.E.approx} 
\end{equation}
This result is exact for all $P$ that are a power of $2$ and serves
as an excellent lower bound for $P$'s that are not.  

For periods that are not a power of $2$, the minimum-$\TI$ words are
those for which the distribution over subwords is as uniform as
possible.  For example, for $P=10$ there are two prime words with
minimum $\TI$:  $0000101111$ and $0001011101$.  Each has $\TI_{\rm
min}$ of $7.39483$.  Note that for each word, the distribution over
subwords of length $1$ is uniform, while the distribution over words
of length $2$ is as uniform as can be, given that there are $10$
occurrences of subwords of length $2$; for each, two length-$2$
subwords occur with a frequency of $3/10$ and two occur with a
frequency of $2/10$.


We have obtained an analytic expression for $\TI_{\rm min}$ for
sequences of general period $P$.  The main idea, as stated above, is to
distribute the frequencies of the subwords as uniformly as possible.
How uniform this distribution is depends on how $2^L$ (the number of
subwords of length-$L$) divides into the period $P$.  Following this
line of reasoning, we obtain the following results.  Let
\begin{equation}
  n_p(L) \, \equiv \, \bigg \lfloor \, \frac{P}{2^L} \, \bigg \rfloor \;
\end{equation}
and
\begin{eqnarray}
N_{\rm odd}(L) \, &\equiv &\, P \;{\rm mod}\; 2^L ~, \\
N_{\rm even}(L) \, &\equiv& \, 2^L - N_{\rm odd}(L) ~.
\end{eqnarray}
Here, $\lfloor x \rfloor$ denotes the {\em floor} of $x$ --- the largest
integer smaller than $x$.  Similarly, $\lceil x \rceil$ is the
{\em ceiling} of $x$ --- the smallest integer larger than $x$. 

The $H(L)$ curve converges to $\EE = \log_2 P$ by length $L^*$, where 
\begin{equation}
 L^* \, \equiv \,  \bigg \lceil \frac{\log_2 P}{2} \bigg \rceil \, +
 \, 1 \;.
\end{equation}
Then, we find that the transient information is given by:
\begin{equation}
\TI_{\rm min} \, = \, (L^* + 1)\log_2 P - \phi \;,
\label{Tmin.exact}
\end{equation}
where
\begin{eqnarray}
\phi \, &=& \, \sum_{n=1}^{L^*} \left[  -N_{\rm
odd}\frac{n_p(L)+1}{n}\log_2 \left( \frac{n_p(L) +1}{n} \right)
\right. \, 
\nonumber \\
& & \hspace{1cm} - \, \left.  N_{\rm even}(L) \frac{n_p(L)}{n} \log_2
\left( \frac{n_p(L)}{n} \right) \right] \;.
\end{eqnarray}
Values obtained using Eq.~(\ref{Tmin.exact}) agree with the exact
results of Table \ref{BigTable}.

\subsection{Average $\TI$}

In Fig.~\ref{T.max.min} we have also plotted $\langle \TI \rangle$,
the average value of the transient information, as a function of the
period $P$.  These average values were also given in Table
\ref{BigTable}.  The average values appear to be well approximated by 
\begin{equation}
  \langle \TI \rangle \, = \, \frac{1}{2} \log_2^2 P + \log_2 P \;. 
\label{Tave.fit}
\end{equation}
This is an empirical fit.  Note that $\langle \TI \rangle$ and
$\TI_{\rm min}$ of Eq.~(\ref{Tmin.approx}) are the same to leading
order in $P$.  This is not surprising, given the asymmetry in the
distribution of $\TI$ values evident in Fig.~\ref{THistogram}.

\subsection{Minimum-$\ST$ Sequences}

For a given period $P$, we found that the sequences with the
minimum transient information $\TI_{\rm min}$ are the same as the
sequences with the minimum synchronization time $\ST_{\rm min}$.  As
noted above, these sequences are those for which the distribution of
subwords is most nearly uniform at each subword length.  Using this
observation, it is possible to derive an analytic expression for
$\ST_{\rm min}$ among the sequences of period $P$. 

As discussed above when considering the minimum-$\TI$ word, if $P =
2^M$, where $M$ is an integer, then the minimum-$\tau$ word is such
that all subwords appear with equiprobability.  In particular, each
subword of length $\log_2 P$ occurs exactly once.  Thus, regardless of
the phase in which one begins observing such a sequence, there is no
ambiguity about the phase of the sequence after making $\log_2 P$
observations.  For observations shorter than $\log_2 P$, there will
always be some uncertainty about the phase of the process.  Thus, the
synchronization time is:   
\begin{equation}
\ST_{\rm min} \, = \, \log_2 P \, = \, \EE \;.
\label{tau.min.power.2}
\end{equation}
This result is exact when $P = 2^M$ and is an excellent approximation
for other periods.

As mentioned above, $\EE$ is the total phase information stored in the
periodic source; it is simply the Shannon entropy of the $P$ possible
phases.  Recall that the entropy of a random variable sets a lower
bound on the number of yes-no questions needed, on average, determine
the value of that variable.  The synchronization time $\tau$ is
equivalent to this average number of yes-no questions --- measuring a
binary symbol entails asking a yes-no question of the source.
Eq.~(\ref{tau.min.power.2}) thus shows that the periodic word with the
minimum $\tau$ saturates the lower bound set by the entropy. 

As was the case for $\TI_{\rm min}$, one can derive an expression for
$\ST_{\rm min}$ for periods that are not a power of $2$.  As with the
minimum-$\TI$ word, in the minimum-$\tau$ word the distribution over
the subwords is as uniform as possible.  In the case
where $P = 2^M + 1$, $P-2$ phases will synchronize in $M$
observations.  The remaining two phases, though, require an
additional observation to synchronize; an additional symbol must be
seen to distinguish between these two phases.  This was the case, for
example, for the period-$5$ case considered in Fig.~\ref{Per5.Tree}.
Here, $P = 5 = 2^2 + 1$ and $M = 2$.   Three of the phases synchronize
after two observations, while two phases synchronize after three
observations.  

Generalizing this observation to $P = 2^M + N$, we obtain the
following result: 
\begin{equation}
  \ST_{\rm min} \, = \, \frac{1}{P} \left[\,  2N(M\!+\!1) +
(P \! - \!  2N)M \, \right] \;, 
\label{tau.min}
\end{equation}
where
\begin{equation}
   M \, = \, \bigg \lfloor \log_2 P \bigg \rfloor
\label{M.def}
\end{equation}
and
\begin{equation}
  N = P \; {\rm mod}\; 2^M \;. 
\end{equation}
Eq.~(\ref{tau.min}) can also be written as \cite[seq.~A061717]{Sloa}:
\begin{equation}
   \ST_{\rm min} \, = \,  \frac{1}{P}\bigg \lceil \log_2 P^P \bigg
   \rceil \;.
\label{tau.min.exact}
\end{equation}
The minimum-$\ST$ values given by Eq.~(\ref{tau.min}) agree with the
exact numerical results shown in Table \ref{BigTable}. For large $P$,
Eq.~(\ref{tau.min.exact}) scales as $\log_2 P$, in agreement with
Eq.~(\ref{tau.min.power.2}).   

\subsection{Maximum-$\ST$ Sequence} 

Recall that the maximum-$\TI$ sequence consisted of $P\!-\!1$ $0$'s
followed by one $1$.  For example, the $\TI_{\rm max}$ sequence is
$000000001$ for $P=9$.  However,
this is {\em not} the maximum-$\ST$ sequence.  The sequence
$000000001$ has an $H(L)$ that grows very slowly, leading to its large
$\TI$.  However, once one sees a $1$, one is synchronized.  Thus, it
is possible for an observer to get lucky, see a $1$ after one or two
observations and be synchronized quickly.  As a result, the
expected synchronization time for this sequence is relatively low,
only $\ST = 4.88889$. 

In contrast, the sequence with the largest synchronization time for
period $9$ is $000010001$ and has $\ST = 6.2222$.  This sequence has
the maximum $\ST$ due the presence of an additional $1$ which prevents
an observer from synchronizing after a single observation; this leads
to a larger $\ST$.  

The maximum-$\ST$ word for periods $4$-$16$ are given in Table
\ref{Tau_Max_Table}.  Note that a maximum-$\ST$ word need not be
unique, although we find that it is unique for all periods examined,
except for $P=6$. 

For $P$ odd, we noticed the maximum-$\ST$ word takes on a particularly
simple form: two $1$'s separated by $(P-3)/2$ zeros.  This allows us
to determine a simple expression for $\ST_{\rm max}$ in this case. We
find: 
\begin{equation}
    \ST_{\rm max} \, = \, \frac{3}{4}P - \frac{1}{2} - \frac{1}{4P}
    \,, \;\;P \; {\rm odd}\; .
\label{Tmax.odd}
\end{equation}

If the period is even, it is also possible to write down an expression
for $\ST_{\rm max}$.  If $P$ is divisible by $4$, then
\begin{equation}
      \ST_{\rm max} \, = \, \frac{3}{4}P - \frac{1}{2} - \frac{1}{P}
      \, , \;\; {\rm P \; divisible \; by \;} 4\;. 
\label{tau.max.4}
\end{equation}
However, if $P$ is divisible by $2$, but not by $4$, then:
\begin{equation}
  \ST_{\rm max} \, = \, \frac{3}{4}P - \frac{1}{2} - \frac{4}{P} \, ,
   \;\; {\rm P \; divisible \; by \;} 2 \; {\rm and \; not \; } 4 \;.
\label{tau.max.2}
\end{equation}
Note that, in all cases, to leading order in $P$, $\ST_{\rm max}$
grows linearly with $P$:
\begin{equation}
   \ST_{\rm max} \, \sim \, \frac{3}{4} P \;.
\label{Tau.Scaling.Form}
\end{equation}
Also, in all cases the preceding results agree with the exact values
obtained by enumeration and shown in Tables \ref{BigTable} and
\ref{Tau_Max_Table}.  Eqs.~(\ref{Tmax.odd}) through (\ref{tau.max.2})
were obtained by noting various regularities in the enumeration data
and not via a first-principles calculation of the subword statistics.  

It is also possible to calculate the transient information $\TI$ for
the word that maximizes $\tau$ for the case in which $P$ is odd.
Recall that the $\tau_{\rm max}$ word in this case consisted of two
$1$'s separated by $(P-3)/2$ zeros.  A brute force, direct counting
approach yields, for large $P$:
\begin{equation}
  \TI_{\rm max \; \tau} \, \sim \, \frac{1}{4}P \EE + \frac{P}{2} \;. 
\label{T.for.Tmin.odd}
\end{equation}

\begin{figure}[tbp]
\epsfxsize=3.2in
\begin{center}
\leavevmode
\epsffile{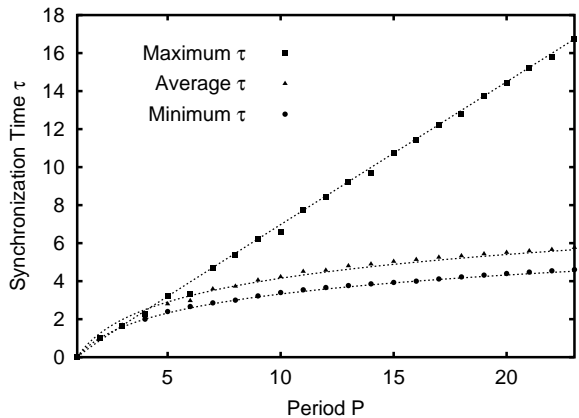}
\end{center}
\caption{The minimum, maximum, and average synchronization time
plotted as a function of period.  The solid points are the exact results
from the enumeration; the dashed lines are the approximations developed
in the text.  The approximation for $\ST_{\rm max}$ is that of
Eq.~(\ref{Tmax.odd}), which is exact for odd periods.  
The approximation for $\ST_{\rm min}$ is that of
Eq.~(\ref{tau.min.power.2}), which is exact for periods that are a
power of $2$.  The approximation for $\langle \TI \rangle$ is the
empirical fit of Eq.~(\ref{tau.ave.fit}).} 
\label{min.max.tau}
\end{figure}

\subsection{Average Synchronization Time $\langle \ST \rangle$}

Figure \ref{min.max.tau} also shows $\langle \ST \rangle$, the
average synchronization time at a given period, as a function of the
period.  The dashed line in this figure is an empirical fit:
\begin{equation}
  \langle \ST \rangle \, \approx \, \frac{5}{4} \log_2 P \;.
\label{tau.ave.fit}
\end{equation}
As was the case with the transient information, the average and
minimum synchronization times appear to grow asymptotically at the
same rate. 

\section{Relationships between $\ST$ and $\TI$}
\label{Relationships.Section}

Throughout the foregoing we have argued that there are important
structural differences between distinct periodic sequences at a given 
period.  Moreover, the structural properties captured $\TI$ and $\ST$,
though both related to synchronization, can be different.  Perhaps the 
differences are not surprising. Based on their definitions these
quantities have different interpretations: $\TI$ measures the total
uncertainty experienced while synchronizing; $\ST$ measures the
expected number of observations needed in order to synchronize.
To clarify the differences and similarities, in this section we
analyze more directly the relationship between $\TI$ and $\ST$.  We
shall see that, while there are relationships between $\TI$ and $\tau$
for the extremal synchronization behaviors, there is a fairly wide
range of $\TI$ and $\tau$ combinations possible among different words
of the same period.  

\begin{table}
\begin{center}
\begin{tabular}{c d l }
 Period & $\ST_{\rm max}$ & $\ST_{\rm max}$ word  \\
\hline
\hline
4   & 2.25  & 0001 \\
5	&3.2	& 00101 \\
6	&3.3333	& {000101,~000001} \\
7	&4.71429 &0001001 \\
8	&5.375	& 00100101	 \\
9	&6.222	& 000010001  \\
10	&6.6 	& 0001001001 \\
11  &7.72727 & 00000100001 \\
12	&8.41667 & 001010010101 \\
13	&9.23077 & 0000001000001 \\
14	&9.71429 & 00001000010001 \\
15	&10.7333 & 000000010000001 \\
16	&11.4375 & 0010101001010101 \\
\end{tabular}
\caption{Maximum synchronization-time words and $\ST_{\rm max}$
  values as a function of period $P$.}
\label{Tau_Max_Table}
\end{center}
%
\begin{center}
\begin{tabular}{c c c }
 Function & Minimum & Maximum  \\
\hline
\hline
 & & \\
$\TI$ & $\frac{1}{2}\log_2^2 P = \frac{1}{2} \EE^2$ 
	& $\frac{P}{2} \log_2 P = \frac{P}{2} \EE$ \vspace{2mm} \\
 $\ST$ & $\log_2 P = \EE $  & $\frac{3}{4} P$ \\
 & & \\
\end{tabular}
\caption{Summary of minimum and maximum $\TI$ and $\ST$ behaviors
expressed as a function of period $P$ and as a function of the excess
entropy $\EE$.}
\label{Min.Max.Summary.Table}
\end{center}
\end{table}

\subsection{Extreme Synchronizations}

We start by summarizing the behaviors of the minimum and maximum values
of the transient information $\TI$ and the synchronization time $\ST$
to leading order in the period $P$ in Table
\ref{Min.Max.Summary.Table}.  Note the enormous range in $\TI$ and
$\ST$ values. For example, if $P=256$, $\TI_{\rm min} \approx 32$,
while $\TI_{\rm max} \approx 1024$.  Similarly, for $P=256$, $\ST_{\rm
min} \approx 8$ while $\ST_{\rm max} \approx 192$.  

Note that for both the minimum and maximum cases, $\TI$ and $\tau$ are
related to each other by:
\begin{equation}
 \TI  \, = \, a \tau \EE \;,
\label{relationship}
\end{equation}
where $a$ is a constant that does not depend on the period $P$.  For
the minimum-($\TI$, $\tau$) case $a = 1/2$ and for the maximum case
$a = 2/3$.  Recall, however, that the maximum-$\TI$ word is not the
same as the maximum-$\tau$ word.  If one uses $\TI_{\rm max}$ and
$\tau_{\rm max \; T}$ from Eq.~(\ref{tau.for.Tmax}) in
Eq.~(\ref{relationship}), one finds $a = 1$.  And, if one uses
$\TI_{\rm  max \; \tau}$ and $\tau_{\rm max}$, one find $a = 1/3$.  

In all cases, note that the ratio of $\TI$ to $\EE$ gives a quantity
proportional to $\tau$.  This ratio may thus be viewed as setting a
characteristic time for synchronization.

\subsection{Nonextreme Synchronizations}

\begin{figure}[tbp]
\epsfxsize=3.2in
\begin{center}
\leavevmode
\epsffile{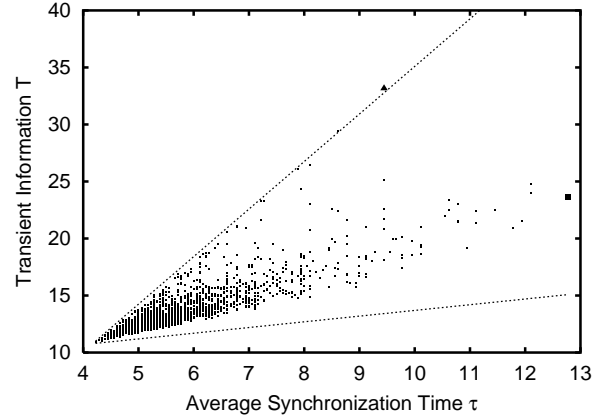}
\end{center}
\caption{A scatter plot of $\TI$ versus $\ST$ for all period-$18$
  sequences.  The upper bound is that of Eq.~(\ref{upper.bound}) and
the lower bound is that of Eq.~(\ref{lower.bound}).
  The solid triangle is the $(\ST, \TI)$ point for the $\TI_{\rm max}$
sequence; the solid square denotes the $\ST_{\rm max}$ sequence.} 
\label{18.scatter}
\end{figure}

It turns out, however, that these simple relations mask a wide
diversity of synchronization behaviors.  Fig.~\ref{18.scatter}, a
scatter plot of $\TI$ and $\ST$ for all period-$18$ sequences, shows
that the relationship between $\TI$ and $\ST$ for individual sequences
is quite a bit more complicated. For example, recall from the previous
section that the maximum-$\TI$ sequence does not correspond with the
maximum-$\ST$ sequence. In the Fig.~\ref{18.scatter}, the maximum-$\TI$
value is shown by a solid triangle and the maximum-$\tau$
value by a solid square.  The minimum-$\TI$ and -$\ST$ sequences are
identical, however.  This corresponds to the single point in the
figure's lower left corner. 

The wide range of points in the scatter plot of Fig.~\ref{18.scatter}
makes it clear that $\TI$ and $\ST$ do indeed measure different
properties: $\TI$ and $\ST$ are not simply rescaled versions of each
other.  Any simple functional relationship is precluded by the diffuse
scatter of points. Said another way, if one lists sequences of a given
period in order of increasing $\TI$, this order will not be the same
as listing the sequences in order of increasing $\ST$. 

Although there is considerable spread evident in Fig.~\ref{18.scatter},
there are bounds limiting the range of $\TI$ values at each $\ST$.

In fact, one can develop an approximate form for the upper bound in
Fig.~\ref{18.scatter} as follows.  Given an arbitrary $\ST$ value and
the period $P$, we are interested in determining the largest possible
$\TI$.  Call this maximum value $\TI_{\rm upper}$.  Recall that the
synchronization time $\tau_{\rm max \; T}$ for the maximum $\TI$ word
scales as $P/2$ for large $P$, as shown in Eq.~(\ref{tau.for.Tmax}).
This lets us write 
\begin{equation}
  \TI_{\rm max} \, = \, \frac{1}{2}P \EE \, = \, \EE \tau_{\rm max \;T} \;.
\label{upper.bound.relationship}  
\end{equation}
To maximize $\TI$ for a given $\tau$, assume that the given $\tau$ is
that which maximizes $\TI$.  Then, using the above equation, we see
that  $\TI_{\rm upper} \propto \EE \tau$.  That is, the slope of the
linear upper bound is $\EE$.  
To get the full equation of the line, we require that it go through
the minimum-$(\ST, \TI)$ point on the lower left hand corner of
Fig.~\ref{18.scatter}.  Using Eqs.~(\ref{Tmin.approx}) 
and (\ref{tau.min.power.2}) for the values of $\TI_{\rm min}$ and
$\ST_{\rm min}$, respectively, we then obtain
\begin{equation}
  \TI_{\rm upper} \, \approx \, \EE\, [ \ST - \frac{1}{2} (\EE - 1 ) ]\;. 
\label{upper.bound}
\end{equation}
This upper bound is plotted in Fig.~\ref{18.scatter}, with
$\EE = \l2 18 \approx 4.1699 \ldots$.  This upper bound is only
approximate because the minimum-($\tau, \TI$) values used in its
calculation are the asymptotic values for large $P$.  The bound is
also approximate because the relationship in
Eq.~(\ref{upper.bound.relationship}), which is true for the
maximum-$\TI$ sequence (the solid triangle in the figure), is assumed
to hold for all sequences.  Despite these approximations, the linear
upper bound appears to fit the data quite well, as shown in
Fig.~\ref{18.scatter}. 

A similar line of reasoning allows us to form a lower bound
$\TI_{\rm lower}$ for the smallest possible $\TI$, given a value of
$\tau$ and the period $P$.  From Eqs.~(\ref{Tmin.E.approx}) and
(\ref{tau.min.power.2}), for large $P$ we know that
$\TI_{\rm min} \approx \frac{1}{2}E^2$ and
$\tau_{\rm min} \approx \EE$.  This gives us
\begin{equation}
  \TI_{\rm min} \, = \, \frac{1}{2} \EE \tau_{\rm min} \;.
\end{equation}
To minimize $\TI$, we assume that the given $\tau$ is the minimum
$\tau$. This tells us that the slope of a linear lower bound is
$\frac{1}{2} \EE$.  Requiring the line to go through the minimum
$\tau, \TI$ point yields
\begin{equation}
  \TI_{\rm lower} \, = \, \frac{1}{2} \tau + \frac{1}{2}E^2 \;. 
\label{lower.bound}
\end{equation}
This is only an approximate bound, for the same reasons that $\TI_{\rm
upper}$ is approximate.  Eq.~(\ref{lower.bound}) is plotted in
Fig.~\ref{18.scatter}.  It is indeed a lower bound, but the bound is
not very tight.

\section{Discussion and Conclusion}
\label{Conclusion}

We began by introducing two measures of how difficult it is to
synchronize to a sequence:  the transient information $\TI$, an
information-theoretic measure of the total uncertainty experienced by
an observer during the synchronization process; and the synchronization
time $\ST$, the number of symbols an observer expects to measure before
it is synchronized.  We exactly calculated $\TI$ and $\ST$ for all
synchronization-distinct periodic sequences up to and including
period $23$.  We also derived a handful of analytic expressions and
approximations for the minimum, maximum, and average $\TI$ and $\ST$
as a function of period.  

Our results show that there are many structural differences between
periodic sequences, even within sequences of the {\em same period}.
These differences are simply not captured by commonly used information
theoretic measures, such as the entropy rate and the excess entropy.  
Are all periodic sequences of a given period the same from an
information-theoretic standpoint? The answer we gave here is ``no'';
the recently introduced transient information, in fact, captures the
information-theoretic differences between sequences of a given period.
Are there structural differences between different periodic sequences
of the same period? We showed that the answer is ``yes'' and argued
that the differences are significant. Are there distinct classes of
periodic sequence? Again, we provided a positive answer, partly by
contrasting properties captured by the transient information with a
sequence's group theoretic properties and its synchronization time.
The latter measures the average number of observations needed to
synchronize to the sequence and, for a given sequence at a given
period, this number was neither the same nor simply proportional to
the transient information. Nonetheless, at a coarse level, the
leading-order approximations we developed provide simple and direct
links between the three different concepts of phase memory $\EE$,
synchronization time $\ST$, and the total uncertainty experienced
during synchronization $\TI$. 

There are a number of areas in which our results may find application.
For one, recognizing the phases of long periodic sequences of the type
investigated here is a key technology for current and future large-scale
satellite-to-satellite communication systems. These multiagent systems
require extremely accurate and robust satellite-to-satellite
synchronization \cite{Mass00a} generally for coordination in signal
processing and navigation and specifically for estimating
inter-satellite signal delays and distances. Synchronization between
two satellites is effected by one transmitting a very long ($P \sim
1,000$) and known binary periodic sequence. The receiver then infers
the phase, using a hierarchical correlation algorithm adapted to
account for noise corruption of individual symbols during
transmission. It turns out that the long periodic {\em acquisition
sequences} are aperiodic sequences formed from relatively ``prime''
short aperiodic sequences. In fact, the synchronization properties 
of the component sequences and the composite acquisition sequence
determine bounds on the computational effort and noise robustness that
can be achieved by the receiver's detection system. We conjecture that
the transient information of the acquisition sequence is a key
parameter determining these properties. 

More generally, these results bear on any situation in which an
intelligent agent needs to learn the phase of a periodic component
in its environment's behavior; cf. Ref. \cite{Crut01b}. This arises
in multiagent systems if an agent needs to know where it is in a
physical environment that varies periodically in space or time,
if it needs to adjust its behavior in response to the repetitive
behavior of other agents, or if a collective decision requires
coordinated information processing.

Finally, we mention several open questions and directions for future
work.  First, do these results hold if the synchronization process is
noisy and bits flip occasionally?  They should extend relatively
directly to a class of ``noisy-periodic'' processes in which one or
several symbols in a periodic sequence are random. However, can these
results be extended to the full class of finite-memory sources? For a
further discussion of the synchronization scenario in a setting not
restricted to periodic sequences see Refs.~\cite{Crut01a} and \cite{Crut01b}.

Second, one may view those sequences at a given period with the same
$\TI$ as forming an equivalence class.  Recall that distinct
periodic sequences were defined by the group $S_k \times C_P$. 
These are the equivalence classes of zero entropy rate, finite excess
entropy sequences. But what new algebra or symmetry group of periodic
sequences 
characterizes those sequences that are equally hard to synchronize to,
as measured by $\TI$ or $\tau$?  If we can characterize identical
$\TI$ or $\tau$ sequences in terms of a symmetry group or algebra, it
should be possible to obtain an analytic expression for the number of
distinct $\TI$ or $\tau$ values at a given period. 

Third, as noted above, Titsworth \cite{Tits64b} pointed out that
Eq.~(\ref{periodic.words}) gives the number of distinct finite-state
machines needed to generate all sequences of a given prime period.
The picture is that starting in a different state of such a machine gives 
the cyclic permutations, while exchanging the output symbols gives
the permutation $0 \leftrightarrow 1$. Titsworth does not comment
on the structure of the synchronizing states of these machines.
To account for synchronization one wants each finite-state machine
to have a unique start state that corresponds to the condition of an
observer not knowing in which phase a periodic sequence starts. It
turns out that $\epsilon$-machines \cite{Crut89,Crut92c,Shal01a}
capture this synchronizing structure in their transient states, and
these in turn determine $\TI$.  Moreover, this is true for both
periodic and random sources. Based on these observations we conjecture
that the number of distinct $\TI$ values for a given period is the
number of distinct $\epsilon$-machines needed to recognize all
periodic sequences of a given prime period.  That is, we conjecture
that those sequences with the same $\TI$ value have the same transient
$\epsilon$-machine structure, up to edge relabeling.

In summary, we analyzed what is arguably the simplest class of
sequences:  zero entropy rate, finite-period sequences. The sequences  
represent behaviors that are often ignored in many fields as being too
simple. As we demonstrated, however, their structural complexity
and synchronization properties are rich and subtle. Within a given
period, there are large differences in structural complexity, as
measured by the transient information $\TI$ and the synchronization
time $\ST$. Thus, even simple periodic systems are surprisingly
complex.  

\section*{Acknowledgments}

The authors thank James Massey for sending his lecture notes.
This work was supported at the Santa Fe Institute under the
Computation, Dynamics, and Inference Program via SFI's core grants
from the National Science and MacArthur Foundations. Direct support
was provided from DARPA contract F30602-00-2-0583. DPF thanks the
Department of Physics and Astronomy at the University of Maine for
its hospitality. 


\bibliography{spin}

\end{multicols}

\end{document}